# To Click or not to Click? The Role of Contextualized and User-Centric Web Snippets

Nikos Zotos
Computer Engineering Dept
Patras University, Greece
zotosn@ceid.upatras.gr

Paraskevi Tzekou
Computer Engineering Dept
Patras University, Greece
tzekou@ceid.upatras.gr

George Tsatsaronis
Athens University of Economics
and Business, Greece
gbt@aueb.gr

Lefteris Kozanidis
Computer Engineering Dept
Patras University, Greece
kozanid@ceid.upatras.gr

Sofia Stamou
Computer Engineering Dept
Patras University, Greece
stamou@ceid.upatras.gr

Iraklis Varlamis
Athens University of Economics and
Business, Greece
varlamis@aueb.gr

## ABSTRACT
When searching the web, it is often possible that there are too many results available for ambiguous queries. Text snippets, extracted from the retrieved pages, are an indicator of the pages' usefulness to the query intention and can be used to focus the scope of search results. In this paper, we propose a novel method for automatically extracting web page snippets that are highly relevant to the query intention and expressive of the pages' entire content. We show that the usage of semantics, as a basis for focused retrieval, produces high quality text snippet suggestions. The snippets delivered by our method are significantly better in terms of retrieval performance compared to those derived using the pages' statistical content. Furthermore, our study suggests that semantically-driven snippet generation can also be used to augment traditional passage retrieval algorithms based on word overlap or statistical weights, since they typically differ in coverage and produce different results. User clicks on the query relevant snippets can be used to refine the query results and promote the most comprehensive among the relevant documents.

## Categories and Subject Descriptors
H.3.3 [**Information Search and retrieval**]: Selection Process, Information Filtering; H.3.1 [**Content Analysis and Indexing**]: Linguistic Processing; H.3.4 [**Systems and Software**]: Performance Evaluation (efficiency and effectiveness).

## 1. INTRODUCTION
The advent of the web has brought people closer to information than ever before. Web search engines are the most popular tool for finding useful information about a subject of interest.

What makes search engines popular is the straightforward and natural way via which people interact with them. In particular, people submit their requests as natural language queries and they receive in response a list of URLs that point to pages which relate to the information sought. Retrieved results are ordered in a way that reflects the pages' importance or relevance to a given query. Despite, the engines' usability and friendliness, people are often-times lost in the information provided to them, simply because the results that they receive in response to some query comprise of long URL lists. To fill this void, search engines accompany retrieved URLs with snippets of text, which are extracted either from the description meta-tag, or from specific tags inside the text (i.e. title or headings).

A snippet is a set of (usually) contiguous text, typically in the size of a paragraph, which offers a glimpse to the retrieved page's content. Snippets are extracted from a page in order to help people decide whether the page suits their information interest or not. Depending on their decisions, users might access the pages' contents simply by clicking on their URLs (retrieved by the engine) or ignore them and proceed with the next bunch of results.

Most up-to-date web snippet generation approaches extract text passages[1] with keyword similarity to the query, using statistical methods. For instance, Google's snippet extraction algorithm [1] uses a sliding window of 15 terms (or 100 characters) over the retrieved document to generate text fragments in which it looks for query keywords. The two passages that show up first in the text are merged to produce the final snippet. However, statistically generated snippets are rough indicators of the query terms co-occurring context but, they lack coherence and do not communicate anything about the semantics of the text from which these are extracted. Therefore, they are not of much help to the user, who must decide whether to click on a URL or not.

Evidently, if we could equip search engines with a powerful mechanism that generates self-descriptive and document expressive text snippets, we could save a lot of time for online information seekers. That is, if we provide users with that piece of text from a page that is the most relevant to their search intention and which is also the most representative extract from the page, we may assist them decide whether to click on the page or not.

---
[1] We use the terms snippet and passage interchangeably to denote the selection of small size text from the full content of a document.

In this paper, we propose a snippet selection technique, which relies on the implicit query semantics rather than the query terms and on the snippets semantic information rather that on the statistical distribution of terms within the text. Our technique focuses on selecting *coherent*, *query-relevant* and *expressive* text fragments, which are delivered to the user and which enable the latter perform focused web searches. At a high level our method proceeds as follows:

- It takes as input a query and uses a number of semantic resources (thesauri, ontologies, etc.) in order to assist the user in determining the query intention. This practically translates into offering the user the means to annotate search terms with the appropriate sense (always specified in the query context).
- Given the disambiguated query intention and a set of results that correlate to the underlying intention, it identifies within the text of a page, the fragment that is the most relevant to the semantics of the query.
- Query-relevant text snippets are then evaluated in terms of their lexical elements' coherence, their importance to the semantics of the entire page and their closeness to the query intention.
- Snippets that exhibit the strongest correlation to both the query and the page semantics are presented to the user.

After applying our snippet selection approach to a number of searches, we conclude that retrieved snippets determined by the semantic correlation between snippets and queries yield improved accuracy compared to the snippets that are determined by using only the statistical distribution of query keywords in the pages' snippets. In brief, the contributions of this article are as follows:

- A measure of the snippet's closeness to the query intention (usefulness). In our work, a useful snippet is the text fragment in a retrieved page that exhibits the greatest terminological overlap to the query keywords and which is also semantically closest to the query intention.
- A measure of the importance and representativeness of a snippet against the entire document from which it derived. Our measure adheres to the semantic cohesion principle and aims at identifying the query focus in the search results.
- A combination of the above measures, in order to assist the user in performing comprehensive and focused web searchers in two ways: with and without clicking on the retrieved results. Without clicking on the results, the user can view the particular text fragment in the page that best matches her search intention. By clicking on a snippet, the user's focus is directly driven to the exact text fragment that contains relevant information to her search intention. In particular, query-relevant text fragments appear highlighted so that the user gets instantly the information that she wants without the need to go through the contents of a possibly long document.

The paper is organized as follows. We begin our discussion with a detailed description of our semantically-driven approach in snippets' selection. Then in Section 3, we experimentally study the effectiveness that our snippet selection approach has in focused retrieval and we discus obtained results. In Section 4 we review related work and we conclude the paper in Section 5.

## 2. MINING QUERY-RELEVANT AND TEXT-EXPRESSIVE SNIPPETS

It is common knowledge that web users decide on which results to click based on very little information. Typically, in the web search paradigm, information seekers rely on the retrieved page's title, URL and text snippet that contains their search keywords to infer whether the page is of interest to their search pursuit or not.

Although, the titles given to web pages are greatly representative of their content, the text snippets of the search results might oftentimes be misleading and communicate incomplete information about the pages' semantic content. This is essentially because titles are created manually, whereas web snippets are automatically generated by the search engine modules on the sole ground that they contain the query keywords.

Evidently, decisions based on little information are susceptible to be bad decisions. A bad decision is encountered when the user clicks on a link misguided by a title or a text snippet, which is of little relevance to the linked page's contents. In an analogous manner, a bad decision might be when the user decides not to click on the link to a *good* page simply because the text snippet of the page is poor or seems unrelated to her query intention.

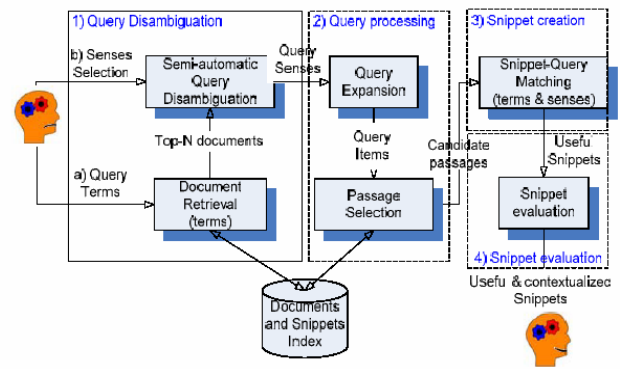

**Figure 1. Snippet selection process**

In this section, we present our approach towards the automatic extraction of query-relevant and document-expressive snippets, in the hope of assisting web information seekers make informative decisions about whether to click on a retrieved result or not. The basic steps of our approach, as depicted in, are:

1) Disambiguation of the query intention, with automatic, semi-automatic or completely manual methods.
2) Semantic similarity matching between query and text passages using both terms and implied concepts (candidate passages).
3) Creation of query-similar snippets from the document (useful snippets).
4) Evaluation of the selected snippet's expressiveness to the document contents.
5) Presentation of the query-relevant and text-expressive snippets to the user.

We begin our discussion, with a brief description of our approach towards the identification of the query intention (step 1). Then, in Section 2.2 we describe our semantically-driven approach for extracting candidate text nuggets from a query matching page (step 2) and selecting those that are semantically closest to the query intention (step 3). In Section 2.3, we introduce a novel method for evaluating how expressive or else representative is a query-relevant text fragment to the entire content of the page from which it derived (step 4). Finally, in Section 2.4, we discuss how we can put together the derived information about the text nugget that is the most useful to the query intention and also expressive of the document's content in order to assist web users perform focused web searches.

## 2.1 Identifying the Query Intention

A number of studies have shown that a vast majority of queries to search engines are short and under-specified [13]. Moreover, short keyword queries are inherently ambiguous in the sense that the same query might intend the retrieval of distinct information sources. Although, the problem of query sense detection is not new, nevertheless the challenge of deciphering the intention of a query still remains.

In our work, we attempt the semi-automatic identification of the query intentions based on the semantic analysis of the query matching pages [14]. In particular, we rely on the top N (N=20) pages retrieved for a query, we parse them to remove html markup, we tokenize, POS-tag them, remove their stop-words and we utilize them as a small Web corpus of query co-occurrence data against which query sense resolution is attempted. The first step towards query sense disambiguation concerns the mapping of all content terms[2] inside every page to WordNet [16] nodes. The corresponding query senses that relate (in WordNet) to any of the senses of the page's content terms are candidate senses for describing the query intention.

For instance, assume that query $q$ has 4 senses in WordNet, say $s_1$, $s_2$, $s_3$ and $s_4$, and that a query matching page $P$ has a number of terms $t_1, t_2, …, t_n$ with senses $t_1\{s_1, s_2\}, t_2\{s_1,s_2,s_3\}, …. t_n\{s_1,s_2\}$. To identify which of the 4 query senses is attributed to $q$ in the contents of $P$, we examine which senses of $q$ relate in WordNet to any of the senses of $t_1,t_2,…t_n$. We then take the query senses that relate to any of the page terms' senses and present them to the user in order to select which of the displayed senses is the most suitable for describing her information need.

In particular, assuming that query sense $s_1$ relates to some sense of $t_1$, query sense $s_3$ relates to some sense of $t_1$ and query senses $s_2$ and $s_4$ do not relate to any of the senses of the terms in $P$, our approach picks the query senses $s_1$ and $s_3$ and displays them to the user as candidate senses for describing the query intention in the context of $P$. The user implicitly indicates the intention of the query, by picking among the candidate concepts, those that she deems the most suitable to express her query semantics. By relying on the user for the final selection of a suitable query sense, we ensure that the query intention is accurately disambiguated and therefore it can successfully participate in the snippet selection process.

Before we proceed with the description of how the identified query sense participates in the snippet selection process, we should stress that our method on snippet selection is not bound to a particular query sense resolution method. Consequently, it can be fruitfully combined with any query disambiguation technique that one would like to use. For an overview of the different similarity metrics employed for word sense resolution see [37].

## 2.2 Semantic Selection of the Best Snippet for a Query

Having identified the query semantics or else the query intention, we now turn our interest to the description of how this knowledge can be exploited in the selection of those document fragments that are semantically closest to the query intention.

The process begins with the expansion of the disambiguated set of query terms with their synonyms in WordNet. This ensures that text fragments containing terms that are semantically equivalent but superficially distinct to the query terms, are not neglected in the snippet selection process. The snippet selection that follows finds all the appearances of the original query terms and their synonyms (query items in) within the retrieved page. Upon identification of query matching items in the page's text, we define a window size of 20 words (see [17]) around the identified query items and we extract all the passages that contain any of the items of the expanded query set. All the extracted passages are candidate snippets with respect to the considered query.

To identify within a query relevant page those text snippets that better match the query intention, our method combines (i) the ***terminological overlap*** (expressed by the relevance measure) and (ii) the ***semantic correlation*** (expressed by the quality measure) between the query and snippet sets of concepts.

The ***terminological overlap*** between the query and a snippet is, in rough terms, the intersection of the two item sets; given that all snippets have similar size and that the items in both sets have been mapped to WordNet nodes. The normalized terminological overlap between a query and a passage, which is determined by the fraction of the passage's terms that have a semantic relation[3] in WordNet to the query sense, indicates the ***relevance*** that a passage has to the query intention and it is formally given by:

$$\text{Relevance}(q, p) = \frac{\sum_{j=1}^{k} qr \cdot \text{Tf/IDF}(t_j, p)}{qs \cdot \sum_{i=1}^{n} \text{Tf/IDF}(t_i, p)}$$

Where $k$ is the number of terms in passage $p$ that relate to at least one term in the query, $n$ is the total number of terms in the passage, $qr$ is the number of query terms to which the passage term $t_j$ relates (query relevant terms) and $qs$ is the number of terms in the query (query size). Finally, Tf/IDF($t_x$,p) denotes the importance of term $t_x$ in passage $p$ as this is determined by their cosine similarity in the vector space model. Passages containing terms that relate to the sense of the query keywords are deemed to be query relevant. However, this is not sufficient for judging the quality or the usefulness that the candidate passages have to the query intention.

To ensure that only good quality passages will participate in the snippet to be extracted from a query matching page, we semantically correlate the expanded query and the query-relevant passage. The query-passage term similarity metric is based on the Wu and Palmer similarity metric [15], which combines the depth of paired concepts in WordNet and the depth of their least common subsumer (LCS), in order to measure how much information the two concepts share in common. According to Wu and Palmer the similarity between a query term $q_i$ and a passage term $S_k$ is given by:

$$\text{Similarity}(q_i, S_k) = \frac{2 * \text{depth}(\text{LCS}(i,k))}{\text{depth}(i) + \text{depth}(k)}$$

---

[2] Content terms are nouns, proper nouns, verbs, adjectives and adverbs.

[3] Out of all the WordNet relation types, in our work we employ: direct hypernymy, (co-)hyponymy, meronymy and holonymy, as indicative of the query-passage terminological relevance.

The average similarity between the query and the passage items indicates the *semantic correlation* between the two. The query passage semantic correlation values, weighted by the score of their relation type (r) that connects them in WordNet, quantifies the quality of the selected passage. Formally, the *quality* of a passage $S$ containing $n$ terms to some query $q$ containing $m$ terms is:

$$\text{Quality}(S, q) = \frac{1}{n \times m} \sum_{j=1}^{m} \sum_{k=1}^{n} \{[\text{Similarity}(q_j, S_k) \bullet \text{RelationWeight}(r)]\}$$

where, RelationWeights(r) have been experimentally fixed to 1 for synonymy, 0.5 for hypernymy and hyponymy and 0.4 for meronymy and holonymy, based on the relation weight values introduced in [18]. The final step towards the identification of the best text nuggets within a query matching page, is to compute the degree to which a candidate passage makes a useful snippet to the user issuing a query and receiving a list of answers in the form of page URLs and accompanying text fragments. In measuring the usefulness that a candidate snippet has to some query, we rely on the combination of the snippet's relevance and quality to the query intention. Formally, the usefulness of a snippet $S$ to a query $q$ is:

$$\text{Usefulness}(S, q) = \text{Relevance}(q, S) \bullet \text{Quality}(S, q)$$

Following the steps described above, in the simplest case, we select from a query matching page the text passage that exhibits the greatest usefulness value to the query intention, as the best snippet to accompany the page retrieved for that query. In a more sophisticated approach, we could select more than one useful passages and merge them in a coherent and expressive snippet.

## 2.3 Towards Coherent and Expressive Text Snippets

Having presented our approach towards selecting query–relevant text snippets, we now proceed with the qualitative evaluation of our selection. The aim of our evaluation is to ensure that the snippets presented to the user are both coherent and text-expressive. By coherent, we mean that the selected snippet should be well-written and meaningful to the human reader, whereas by text-expressive we mean that the selected snippet should represent the semantics of the entire document in which it appears.

Snippet coherence is important in helping the user infer the potential usefulness of a search result before she actually clicks on that. Snippet expressiveness is important after the user clicks on a snippet, since it guarantees that the snippet is in accordance to the target page. Given that our passage selection method operates upon the semantic matching between the query intention and the snippet terms, the evaluation of a snippet's coherence focuses on semantic rather than syntactic aspects. That is, in our evaluation we measure the degree to which terms within the snippet semantically relate to each other. To evaluate semantic coherence of a selected snippet, we map all its content terms to WordNet nodes. Thereafter, we apply the Wu and Palmer similarity metric (cf. Section 2.2) in order to compute the degree to which snippet terms correlate to each other. Based on the average paired similarity values between snippet terms, we derive the degree of the in-snippet semantic coherence as:

$$\text{Coherence}(S_1) = \frac{1}{n} \sum_{i,j=1}^{n} \arg \max_{w_j} \text{similarity}(w_i, w_j)$$

where Coherence denotes the *in-snippet semantic correlation* of terms $n$ in snippet $S_1$. Since the appropriate senses for words $w_i$ and $w_j$ are not known, our measure selects the senses which maximize Similarity (argmax similarity($w_i$, $w_j$)).

Measuring semantic coherence amounts to quantifying the degree of semantic relatedness between terms within a passage. This way, high in-snippet average similarity values yield semantically coherent passages. Semantic coherence is a valuable indicator towards evaluating the degree to which a selected passage is understandable by the human reader. However, even if a passage is semantically coherent, there is no guarantee that the information it brings is expressive of the entire document content.

Snippet *expressiveness is* the degree to which a selected passage is expressive of the entire document's semantics. For modeling the text-expressiveness of a selected passage we want to compute the terminological overlap and the semantic correlation between the selected passage and the rest of its source text. Our computational model is analogous to the query-snippet usefulness metric with the only difference that in our evaluation we compare passages rather that words.

More specifically, we take all the content terms inside a document (passage content terms included), we map them to their corresponding WordNet nodes and we define the *Expressiveness* of a snippet ($S_1$) in the context of document $D$ as follows:

$$\text{Expressiveness}(S_1, (D - S_1)) = \text{Usefulness}(S_1, (D - S_1))$$

where Usefulness ($S_1$,($D$-$S_1$)) denotes the product of (i) the terminological overlap (i.e. Relevance) between the terms in the selected snippet and the terms in the remaining source document (i.e. D-S1) and (ii) the average semantic correlation between the passage and the remaining text items, weighted by their Relation (r) type.

Based on the above formula, we evaluate the level of expressiveness that a selected passage provides to the semantics of the entire text in which it appears. The expressiveness of a snippet increases with the number of semantically related terms between the snippet and the rest of the text in its source document. The combined application of the snippet coherence and expressiveness metrics gives an indication on the contribution of a snippet in conveying the message of a document retrieved in response to a user query.

## 2.4 Snippet-Driven Focused Retrieval

So far, we have described our technique for selecting, from a query matching document, the fragments that are semantically close to the query intention. Moreover, we have introduced qualitative measures for assessing how comprehensive is the selected snippet to the human reader and how indicative it is of the entire document semantics.

We now turn our attention on how we can put together the criteria of usefulness; semantic coherence and text expressiveness, in order to assist users perform focused web searches. The foremost decision is to **balance the influence of each criterion** in our final decision on the best snippet. In other words, we need to decide whether a query-useful snippet should be valued higher than a semantically coherent or text expressive snippet and vice versa.

Apparently, the weight that could or should be given to each of the individual scores cannot be easily determined and even if this is experimentally fixed to some threshold, it still bears subjectivity as it depends on several factors such as the characteristics of the dataset, the user needs, the nature of the query and many other. Whatever the reasons and whichever the objectives for weighting the individual scores within a single formula, in the course of this study we let the final decision on the user, who can apply her own evaluation criteria for selecting which snippet will be displayed for a retrieved document. An approach is to present multiple snippets from each document in the query results (i.e. the best snippet when accounting only one criterion each time) and consequently exploit user feedback to conclude on how users perceive the contribution that different values have on snippet-driven retrieval performance. Based on the users' implicit feedback on what makes a good snippet, we could determine the most appropriate weighting scheme for each user [23].

Another critical issue, concerns the **visualization of the selected snippets** to the end user. We claim that it would be useful to highlight the query terms and their synonyms inside the selected snippets, so that the users can readily detect their search targets. Moreover, it would be convenient that text passages are clickable and upon their selection they direct the user to the query relevant snippet rather than the beginning of the document. This way, we can take off the user the burden of reading through the entire document until she detects the information that is most relevant to her query intention. The snippet selection process can be enriched by merging together snippets from multiple documents and by presenting the merged snippet to the user as an extended answer to her information interests.

In overall, deciding on what makes a good snippet for a particular user information need is a challenge that leaves ample space for discussion, experimentation and evaluation. Next, we present an experimental study that we conducted in order to validate the contribution of our snippet selection method in focused retrieval performance and we discuss experimental results.

## 3. EXPERIMENTS

To validate the usefulness of our snippet selection algorithm in focused retrieval, we conducted two distinct yet complementary experiments. In one experiment we evaluate the performance of our snippet selection algorithm in delivering query useful snippets, and in the second experiment we evaluate how users perceive the query usefulness, the coherence and the text expressiveness of the passages retrieved by our approach.

### 3.1 Experimental Setup

In our study, we compared our semantically driven passage retrieval algorithm against a baseline passage retrieval algorithm. Building upon the machinery of the previous sections, we automatically disambiguated a set of snippets and measured the improvement of incorporating the Usefulness and the Coherence semantic pieces of information into the text retrieval task against a standard baseline.

More specifically, following a similar experimental framework with the one described in [26] we compared the term TF/IDF vector space retrieval model against a retrieval technique utilizing manually disambiguated queries along with the automatically disambiguated snippets set. In our experiment we exploited existing knowledge on the snippets' relevance to their corresponding queries and we evaluated the Usefulness and the Expressiveness of the passages selected by our algorithm.

To quantitatively evaluate the performance of our passage retrieval algorithm, we have employed the NPL data collection [36] as a testbed. NPL contains 93 experimental queries and a total of 11,429 short documents. Out of all the NPL documents and queries we selected a total of 30 queries and their respective 10,737 relevant documents that we processed as described in Section 2.1 and we indexed them in a local SQL 2005 server. Although, NPL provides a well-structured collection of queries and relevant documents and as such it may not be representative enough of the web data, nevertheless it provides a gold standard collection for running preliminary experiments and evaluate the feasibility of our method. Another reason for employing the above dataset is that NPL documents are quite short (i.e. they contain on average 23 terms) and as such they approximate snippets' size. Moreover, the NPL queries vary in size (i.e. between 2 and 9 terms) and constitute partially formed questions rather than mere keywords. As such they are convenient for a passage retrieval experiment.

In the course of our study, we have semi-automatically annotated each of the 30 experimental queries with an appropriate WordNet sense that represents the query intentions. Moreover, we have annotated every word inside all NPL documents with an appropriate WordNet sense through the exploitation of the Wu and Palmer similarity metric. Based on the selected collection of queries, followed by the given gold standard relevant documents, we evaluated the effectiveness of our snippet selection approach in delivering query useful and text expressive snippets.

### 3.2 Query Useful Passages

This experiment aims at comparing our **semantically-driven passage retrieval algorithm**, which computes a query useful passage based on the semantic correlation between the query and the passage terms, against the baseline generated by the term TF/IDF vector space representation of the snippets and the use of cosine for query to snippet similarity. For our comparison, we measure the efficiency of the two algorithms in delivering query useful snippets, which practically translates into comparing the Relevance and Quality values of the snippets retrieved by each of the algorithms for the respective queries. To enable our comparison, we formulate the NPL collection as follows: We merge all NPL documents together into a huge virtual document. This document can answer all queries in our dataset. Every individual NPL document forms a candidate passage into the virtual document, which can answer each of the experimental queries. Given that we know in advance which passage of the huge document (i.e. the entire NPL collection) answers each query, our evaluation proceeds as follows. We employ the baseline algorithm and our semantically-driven algorithm, which combines the snippets' relevance and quality values (i.e. usefulness) and we give scores for each passage. Furthermore, we combine the baseline query to snippet similarity with the computed semantic similarity when the retrieved document reports a Coherence value larger than the average snippet coherence in the collection. We compare the 3 metrics by drawing the interpolated standard 11 precision-recall point curves.

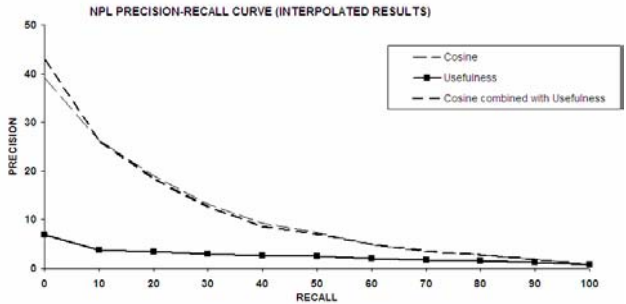

**Figure 2. Performance of query-useful passage retrieval.**
Obtained results indicate that our proposed semantic similarity measures and more specifically the incorporation of usefulness into the query to snippet similarity measure when the snippet coherence is high can aid the text retrieval task. We show that when the usefulness measure for semantically coherent snippets is applied, an improvement of almost up to 3.5% (see Figure 3) can be achieved even by the top three standard precision recall points.

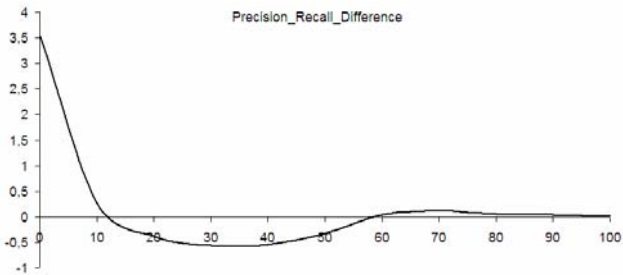

**Figure 3. Performance improvement using semantics.**
Although the retrieval improvement is quite low and therefore might be perceived as statistically insignificant, nevertheless we believe that a 3.5% improvement over a well structured and manually annotated data collection will significantly increase when a semi-structured un-annotated dataset is considered. As such we claim that the improvement our method can achieve in the context of the web retrieval will be much higher that the one obtained in a small and well-balanced document collection.

## 3.3 Impact of Passage Selection Criteria

Having accumulated perceptible evidence on the effectiveness that our semantically-driven passage selection approach has on retrieval performance, we carried out a blind user study in order to evaluate the impact that the different snippet selection criteria have on human judgments. For our study, we employed the 30 NPL queries and their relevant documents to which we applied our snippet selection algorithm three times.

In the first run, we parameterized our algorithm so that it selects from a document the text nugget that is the most useful to the query intention. That is, we applied our Usefulness metric (cf. Section 2.2) to each of the query relevant documents in order to extract from every document the text nugget that is semantically closest to the query semantics. In the second run, we parameterized our algorithm so that it selects from a query relevant document the text fragment that is the most coherent. That is, we applied our semantic Coherence metric (cf. Section 2.3) to each of the query relevant documents in order to extract from every document the piece of text that exhibits the maximum in-snippet semantic correlation. Finally, in the third run, we parameterized our algorithm so that it selects from a query relevant document, the passage that is the most expressive of the documents' semantics. That is, we applied our Expressiveness metric (cf. Section 2.3) to each of the query relevant documents in order to extract from every document the piece of text that most accurately captures the entire document's content.

As a baseline snippet selection technique, we relied on the Alicante passage retrieval algorithm [26] in order to determine a query relevant snippet from each of the query matching documents.

Based on the snippets derived by the baseline, the usefulness-driven, the coherence-driven and the expressiveness-driven selection approaches, we conducted a blind user test, in which we recruited 15 postgraduate students from our university. For our study, we provided our subjects with the list of the 30 sense annotated queries and the snippets selected by each of the algorithms for each of the query relevant documents. The snippets extracted from every query-relevant document were displayed to our users in a random order so as not to convey any information about the criteria under which these were selected. Moreover, in case the same snippet was selected by more that one algorithms, it was presented only once to the user.

We then asked our participants to read all the snippets delivered for each of the queries and indicate for which of the snippets they would like to read the entire source document. In other words, our participants were not informed about the different snippet selection criteria and they were not aware of the fact that all the snippets displayed for a query would direct them to the same document. In contrast, the instructions that were given to them required that: "*Select which of the displayed text fragments do you think will direct you to a document that can successfully answer the search intention of the query?*" Note that the query intention was explicitly communicated to our subjects through the WordNet sense that has been selected for representing the query semantics.

Our participants interacted with a local interface via which we displayed them the annotated experimental queries (one at a time) and the different snippets selected for every query relevant document. For every query, the users viewed at least one snippet (in case all algorithms selected the same passage) and at most four snippets (in case a different passage was selected by each of the algorithms) in a random order. Our subjects indicated their selections by clicking on the snippet that they deemed it would drive them to the most query-relevant document. A user's click on a snippet translates to a vote given by the user for that snippet's success in focusing retrieval results to the query intention.

We recorded the user's clickthrough on the displayed snippets in order to infer the criterion that influences the most human judgments about what makes a snippet successful. In case the user clicked on a snippet that was selected by more than one algorithm, the user's vote was equally attributed to all selection techniques that delivered the particular snippet. Based on the human preferences, we can evaluate to a certain extend how people cast their click decisions to the snippets that they are returned for their search queries. Furthermore, human judgments (reported on Table 1) give us some early intuition about the weights that should be appended to our snippet selection metrics of query usefulness and text expressiveness. The following table reports the number of times each user selected a passage delivered by the baseline, the query usefulness, the semantic coherence and the text expressiveness criteria over all 30 queries examined.

**Table 1. Snippet selection criteria preferred across users**.

| USER | Baseline | Query Usefulness | Semantic Coherence | Text-Expressiveness |
|---|---|---|---|---|
| 1 | 5 | 12 | 9 | 8 |
| 2 | 9 | 17 | 8 | 5 |
| 3 | 6 | 7 | 7 | 10 |
| 4 | 8 | 17 | 9 | 10 |
| 5 | 8 | 13 | 7 | 5 |
| 6 | 11 | 15 | 5 | 6 |
| 7 | 3 | 15 | 7 | 6 |
| 8 | 14 | 14 | 4 | 3 |
| 9 | 9 | 9 | 10 | 7 |
| 10 | 11 | 15 | 5 | 6 |
| 11 | 4 | 11 | 9 | 6 |
| 12 | 7 | 15 | 11 | 8 |
| 13 | 5 | 10 | 11 | 9 |
| 14 | 9 | 18 | 6 | 8 |
| 15 | 6 | 12 | 7 | 5 |

The results of our human survey suggest that semantically derived snippets are valued higher than statistically obtained ones. This is in line with our intuition that passage selection based on the semantic correlation between the passage and query terms yields improved retrieval focus. With respect to the passage semantics, our results demonstrate that the snippet selection criterion that is valued higher by our participants is that of usefulness. This practically implies that what users would like to see in the text fragments accompanying retrieval results are passages that exhibit high semantic and terminological correlation to their query intention. This observation supports the need for more sophisticated approaches towards snippets' selection and indicates that passage retrieval algorithms could be fruitfully explored in this respect. However, our findings relying on a few users and a small number of queries merit further investigation before these are employed towards tuning the weights that should be given to the different metrics employed for snippets' selection.

## 4. RELATED WORK

The role of text snippets or passages in the context of web information retrieval has been studied before. A number of researchers have proposed the exploitation of passages to answer natural language queries [4], [5], [6] and generic queries [3]. Authors in [4] search for single snippet answers to definition questions through the exploitation of lexical rather than semantic patterns. In [5] and [6] the authors exploit WordNet to annotate and consequently answer definition questions. Most of the reported approaches on snippets' exploitation for question-answering rely on some similarity measure in order to derive from a query relevant document, the text fragment that is closest to the query. The relevance/ similarity between the query and the snippet is measured using linguistic [10] (distance in an ontological thesaurus) or statistic [9] (word frequency, proximity or co-occurrence) techniques or a combination of them.

Passage retrieval is a common component to many question answering systems. Currently, there exist several passage retrieval algorithms, such as MITRE [24], bm25 [25], MultiText [34], IBM [28], SiteQ [29], ISI [30]. Recently, [31] quantitatively evaluated the performance that the above passage retrieval algorithms have on question answering. Moreover, passage retrieval approaches have been proposed in the context of web-based question answering [32], [33]. Most of the systems explored in web-based passage retrieval typically perform complex parsing and entity extraction for documents that best match the given queries, which limits the number of web pages that can analyze in detail. Other systems require term weighting for selecting or making the best-matching passages [27] and this requires auxiliary data structures.

Many research works perform post processing of snippets extracted from query results. They either cluster snippets into hierarchies [3], use them to construct ontologies [7], or further expand the snippet collection with relevant information nuggets from a reference corpus [8]. Evaluation of retrieved snippets is performed once again using statistic [35] or linguistic methods [11] and long QA series [12]. Text coherence is a topic that has received much attention in the linguistic literature and a variety of both qualitative and quantitative models have been proposed [19] [20] [21] [22]. Most of existing models incorporate either syntactic or semantics aspects of text coherence.

In our work on passage retrieval, we rely purely on semantic rather that syntactic aspects of both the queries and the documents and we propose a novel evaluation framework which ensures that the passage delivered in response to some query and not merely query relevant but they are also semantically coherent and expressive of the entire document's contents.

## 5. CONCLUSIONS AND FUTURE WORK

In this paper, we have introduced a novel framework for the automatic selection out of a query matching document the text snippet that is the most useful to the query intention. Our approach capitalizes on the notion of semantic correlation between the query keywords and the selected snippet's content as well as on the semantic correlation between the in-snippet terms. We argue that our approach is particularly suited for identifying within the contents of a possibly long document the focus of the query and we introduced a qualitative evaluation scheme for capturing the accuracy in which the selected passage participates in focused web searches. We applied our snippet selection technique to a number of searches that we have performed using synthetic data generated by simulation. Our experiments revealed that our snippet selection method determined by the semantic correlation between the query and the selected text fragment yields increased retrieval performance compared to statistical-based passage retrieval methods.

The snippet selection approach introduced in this paper relies on semantic rather than statistical properties of web documents and it is relatively inexpensive assuming access to a rich semantic resource (such as WordNet). This makes the proposed approach particularly attractive and innovative for the automatic selection and evaluation of focused web snippets. An important future direction lies in the enrichment of our snippet selection model with advanced linguistic knowledge such as co-reference resolution, genre detection or topic distillation. Moreover, it would be interested to experiment with alternative formulas for measuring the correlation between the query keywords and the passage terms, such as the one proposed in [2]. Another possible direction would be to employ a query relevant snippet as a backbone resource for a query refinement technique. Yet a more stimulating challenge concerns the incorporation of user profiles in the snippet selection process in an attempt to deliver personalized text passages. Last but not least, our snippet selection approach could be fruitfully explored in the context of web question-answering and element retrieval systems in the hope of helping the user find the exact information sought in an instant yet effective manner.

## ACKNOWLEDGEMENTS

Authors George Tsatsaronis and Lefetris Kozanidis were funded by the 03ED_850 and 03ED_413 research projects respectively, implemented within the "Reinforcement Programme of Human Research Manpower" (PENED) and co-financed by National and Community Funds (25% from the Greek Ministry of Development-General Secretariat of Research and Technology and 75% from E.U.-European Social Fund). Author Iraklis Varlamis is partially funded by the EPEAEK PYTHAGORAS II research project, financed by the Greek Ministry of Education and Religion.